\documentclass[prl,aps,a4paper,showpacs,showkeys]{revtex4-1}
\usepackage[T1]{fontenc}
\usepackage[latin1]{inputenc}
\usepackage{amsmath}
\usepackage{amssymb}
\usepackage{graphics}
\usepackage{psfrag}
\usepackage{epsfig}
\usepackage{hyperref}
\usepackage{color}

\begin{document}

\title{Phase transitions in the three-state Ising spin-glass model
  with finite connectivity}

\author{R. Erichsen Jr.}  \email{rubem@if.ufrgs.br}
\author{W. K. Theumann} \email{theumann@if.ufrgs.br}
\affiliation{Instituto de Física, Universidade Federal do Rio Grande
  do Sul, Caixa Postal 15051, 91501-970 Porto Alegre, RS, Brazil}

\date{\today}

\begin{abstract}

  The statistical mechanics of a two-state Ising spin-glass model with
  finite random connectivity, in which each site is connected to a
  finite number of other sites, is extended in this work within the
  replica technique to study the phase transitions in the three-state
  Ghatak-Sherrington (or random Blume-Capel) model of a spin glass
  with a crystal field term. The replica symmetry ansatz for the order
  function is expressed in terms of a two-dimensional effective-field
  distribution which is determined numerically by means of a population
  dynamics procedure. Phase diagrams are obtained exhibiting phase
  boundaries which have a reentrance with both a continuous and a genuine
  first-order transition with a discontinuity in the entropy. This may
  be seen as ``inverse freezing'', which has been studied extensively
  lately, as a process either with or without exchange of latent heat.

\end{abstract}



\preprint{IF-UFRGS 2010}

\thispagestyle{empty}

\maketitle

\section*{Introduction}

Numerous problems of frustrated, disordered systems, with extensive
connectivity in which each site is linked to a macroscopic number of
other sites \cite{Ni01} have been studied in the past by means of
the replica technique in mean-field theory \cite{SK76,PMV87}. The
technique has been extended to systems with finite random
connectivity and binary units (spins) in states $\sigma=\pm 1$, in
which each site is linked to a finite number of other sites, in
areas like error correcting codes \cite{MK00,NK01,Ni01}, spin
glasses \cite{VB85,KS87,MP87,WS88,Mo98,MP01}, neural networks
\cite{WC03,PS03,PW04} and small-world lattices \cite{NC04}. The fact
that mean-field theory is exact only for infinite-range interactions
or infinite-dimensional systems has been a challenge for the
understanding of the behavior of more realistic disordered systems,
and a study of the effects of finite connectivity even in mean-field
theory could be a useful improvement.

Three-state spin models of states $\sigma=0, \pm 1$ with random
bonds and finite connectivity could be of interest to condensed
matter physics in view of the phase transitions that already appear,
within mean-field theory, in the random Blume-Emery-Griffith-Capel
(BEGC) model with full connectivity between the spins
\cite{CL04,CL05} and it should also be of interest for information
processing in neural networks \cite{Bo04}. The simplest case of a
fully connected random BEGC model is the three-state spin-glass
model of Ghatak and Sherrington (GS) \cite{GS77,CYS94}, which is a
Blume-Capel (BC) model with random bonds \cite{BC66}. The GS model
with infinite range interactions exhibits both a continuous
transition at high $T$ and a genuine thermodynamic first-order
transition below a tricritical point between a spin-glass and a
paramagnetic phase. The first-order transition appears in a
reentrant part of the phase boundary and it may describe "inverse
freezing" with an exchange of latent heat. This is a reversible
transition, say between a paramagnetic (P) phase and a spin-glass
(SG) phase in which the entropy of the P phase below the transition
is smaller than the entropy of the SG phase and there has been a
recent revival of interest in these transitions \cite{SS03,PL10}
which could explain the behavior of colloidal and polymeric systems,
among others (see ref. \cite{PL10} for a recent summary of
realizations).

A result of the GS model, either in mean-field theory with full
connectivity or obtained by means of numerical simulations for
nearest-neighbor interactions in three dimensions, is that the
tricritical point separating the continuous and the first-order
transition is either above or at the point of reentrance on the
phase boundary implying that inverse freezing appears only as a
first-order transition \cite{CL04,CL05,PL10}.

General results in the form of complete phase diagrams of stationary
states for the GS (or random BC) model with finite connectivity are
still missing and the purpose of this paper is to make progress in
that direction by means of an analytic study extending the SG
replica technique, with the replica-symmetry (RS) ansatz, developed
for disordered Ising systems with finite random connectivity and
binary spins \cite{NC04}. Typical questions one would like to answer
is how the nature of the transition changes with the connectivity,
if there is reentrance behavior of the phase boundaries and if this
takes place either as a continuous or as a first-order transition at
finite temperature.

The procedure on which the present work is based introduces an
order-function \cite{KS87}, in place of an infinite number of order
parameters \cite{VB85}, and uses a representation in terms of a
weighted probability of alignment of the spins involving a
distribution of a two-component local field. One of the components
is associated to a linear form in the spins, while the other
component is associated to the crystal field term. A population
dynamics technique is used to solve numerically a self-consistency
equation for the distribution of the local field that yields the
relevant physical order parameters. Phase diagrams are then obtained
which exhibit either a continuous transition between a P and a SG
phase, for low connectivity, or both a continuous and a first-order
phase transition that merge at a tricritical point with reentrance
behavior characteristic of inverse freezing between those phases,
for increasing finite connectivity.

The outline of the paper is the following. In section 2 we present
the GS model and carry out the replica procedure that yields the
statistical mechanics for the three-state system. In section 3 we
present the results for the distribution of local fields, the
relevant order parameters and the free energy and discuss the phase
diagrams. We conclude in section 4 with a summary and remarks.

\thispagestyle{empty}

\section*{The model and replica procedure}

We consider a system of $N$ interacting three-state Ising spins
$\sigma_i\in\{-1,0,1\}$, $i=1\dots N$, described by the Hamiltonian
\begin{align}
  H=-\frac{1}{c}
  \sum_{i<j}c_{ij}J_{ij}\sigma_i\sigma_j
  +D\sum_i\sigma_i^2\,.
  \label{hamiltonian}
\end{align}
The random variable $c_{ij}\in\{0,1\}$ indicates whether there is a
connection $(c_{ij}=1)$ or not $(c_{ij}=0)$ between a pair of spins
$(i,j)$ and it takes different values for different pairs of spins,
according to the distribution
\begin{align}
  p\left(c_{ij}\right)=\frac{c}{N}\delta_{c_{ij},1}
  +\left(1-\frac{c}{N}\right)\delta_{c_{ij},0}\,\,\,,
  \label{cdistr}
\end{align}
where $c$ (the connectivity) is the average number of connections
per spin which is assumed to remain finite in the thermodynamic
limit $N\rightarrow\infty$, such that $c/N\rightarrow 0$. Thus, the
sites are connected according to a Poisson distribution, and one
makes use of this limit in deriving the statistical mechanics of the
system. There is a set of infinite-range interactions $\{J_{ij}\}$
that will be assumed to be independent, identically distributed,
random variables drawn from a distribution $p(J_{ij})$, to be
specified below, and averages over that distribution of a quantity
$g(J_{ij})$ will be denoted by ${\langle g(J)\rangle}_J$. The
quadratic form in the spins favors the population of the zero state,
if $D>0$, or the states $\pm 1$, if $D<0$. If $D$ is
sufficiently large and negative one retrieves the binary Ising
spin-glass model with finite connectivity and spins
$\sigma_i\in\{-1,1\}$, which is a particular case of the recently
studied small world spin glasses when the nearest neighbor
interaction along the ring is set to zero \cite{NC04}.

Assuming thermal equilibrium at an inverse temperature $\beta=1/T$,
the disorder-averaged free energy per spin is calculated in the
replica procedure as
\begin{align}
  f(\beta)=-\lim_{N\rightarrow\infty}\frac{1}{\beta N}\lim_{n\rightarrow 0}
  \frac{1}{n}\log\langle {Z^n}\rangle\,\,\,,
  \label{freen}
\end{align}
where
\begin{align}
  Z=\sum_{\sigma_{1}\dots \sigma_{N}}{\mathrm e}^{-\beta
    H}
  \label{partition}
\end{align}
is the partition function and the brackets stand for the disorder
average. In the small $c/N$ limit, the disorder-average replicated
partition function becomes, to leading order in $N$,
\begin{align}
  \left\langle Z^n\right\rangle
  =\sum_{\mbox{\boldmath$\sigma$}^1\cdots\mbox{\boldmath$\sigma$}^n}
  \exp\left[-\beta D\sum_{i,\alpha}\left(\sigma_i^\alpha\right)^2
    +\frac{c}{2N}\sum_{i\neq j}\left\langle{\mathrm e}^{\frac{1}
        {c}\beta J\sum_\alpha\sigma_i^\alpha\sigma_j^\alpha}
      -1\right\rangle_J\right] \,\,\,
  \label{rpart3}
\end{align}
where $\alpha=1,\dots, n$ denotes the replica index. Since the
connectivity $c$ is finite, one cannot expand the inner exponential
and follow the standard infinite-connectivity calculation. Instead,
to extract the variables under summation from the inner exponential,
one introduces the identity
\begin{align}
  1=\sum_{\mbox{\boldmath$\sigma$}}\delta_{\mbox{\boldmath$\sigma$}
    \mbox{\boldmath$\sigma$}_i}\equiv\sum_{\mbox{\boldmath$\sigma$}}
  \prod_{\alpha=1}^n\delta_{\sigma^\alpha\sigma_i^\alpha}\,,
  \label{identity1}
\end{align}
where {\boldmath$\sigma$} and $\mbox{\boldmath$\sigma$}_i$ are
$n$-component vectors representing replica states and
$\delta_{\mbox{\boldmath$\sigma$}\mbox{\boldmath$\sigma$}_i}=1$ if
$\mbox{\boldmath$\sigma$}=\mbox{\boldmath$\sigma$}_i$ and zero
otherwise. Thus, we write
\begin{align}
  \left\langle Z^n\right\rangle
  =\sum_{\mbox{\boldmath$\sigma$}^1\cdots\mbox{\boldmath$\sigma$}^n}
  \exp\left[-\beta D\sum_{i,\alpha}\left(\sigma_i^\alpha\right)^2
    +\frac{c}{2N}\sum_{i\neq j}\sum_{\mbox{\boldmath$\sigma$}
      \mbox{\boldmath$\tau$}}\delta_{\mbox{\boldmath$\sigma$}
      \mbox{\boldmath$\sigma$}_i} \delta_{\mbox{\boldmath$\tau$}
      \mbox{\boldmath$\sigma$}_j}\left\langle{\mathrm
        e}^{\frac{1}{c}\beta J\sum_\alpha\sigma^\alpha\tau^\alpha}
      -1\right\rangle_J\right]
  \label{rpart4}
\end{align}
and introduce the order function
$P(\mbox{\boldmath$\sigma$})=(1/N)\sum_i\delta_{\mbox{\boldmath$\sigma$}
  \mbox{\boldmath$\sigma$}_i}$, which represents the fraction of sites
with the replica configuration {\boldmath$\sigma$}, through the
identity
\begin{align}
  1=\int\prod_{\mbox{\boldmath$\sigma$}}dP(\mbox{\boldmath$\sigma$})
  d\hat{P}(\mbox{\boldmath$\sigma$}){\mathrm
    e}^{\sum_{\mbox{\boldmath$\sigma$}}
    \hat{P}(\mbox{\boldmath$\sigma$})\left(P(\mbox{\boldmath$\sigma$})
      -\frac{1}{N}\sum_i\delta_{\mbox{\boldmath$\sigma$}
        \mbox{\boldmath$\sigma$}_i}\right)}\,\,\,,
  \label{identity3}
\end{align}
where $\hat{P}$ is an auxiliary density. Performing the trace and
changing $\hat{P}$ to $N\hat{P}$ the integral
\begin{align}
  \nonumber \left\langle Z^n\right\rangle
  =\int\prod_{\mbox{\boldmath$\sigma$}}dP(\mbox{\boldmath$\sigma$})
  d\hat{P}(\mbox{\boldmath$\sigma$})\exp
  N&\left[\sum_{\mbox{\boldmath$\sigma$}}
    \hat{P}(\mbox{\boldmath$\sigma$})P(\mbox{\boldmath$\sigma$})
    +\log\sum_{\mbox{\boldmath$\sigma$}}{\mathrm
      e}^{-\beta D\sum_\alpha
      \sigma_\alpha^2-\hat{P}(\mbox{\boldmath$\sigma$})}\right.\\
  &\quad\quad\quad\left.+\frac{c}{2}\sum_{\mbox{\boldmath$\sigma$}
      \mbox{\boldmath$\tau$}}P(\mbox{\boldmath$\sigma$})
    P(\mbox{\boldmath$\tau$})\left\langle{\mathrm e}^{\frac{1}{c}\beta
        J\sum_\alpha\sigma_\alpha\tau_\alpha}-1\right\rangle_J\right]
  \,
  \label{rpart6}
\end{align}
can be evaluated by the saddle-point method in the large-$N$ limit,
leading to the extremum
\begin{align}
  \nonumber f(\beta)=-\lim_{n\rightarrow 0}\frac{1}{\beta n}\rm Extr
  &\left[\sum_{\mbox{\boldmath$\sigma$}}
    \hat{P}(\mbox{\boldmath$\sigma$})P(\mbox{\boldmath$\sigma$})
    +\log\sum_{\mbox{\boldmath$\sigma$}}{\mathrm
      e}^{-\beta D\sum_\alpha
      \sigma_\alpha^2-\hat{P}(\mbox{\boldmath$\sigma$})}
  \right.\\
  &\quad\quad\quad\quad\quad\quad\left. +\frac{c}{2}\sum_{\mbox{\boldmath$
        \sigma$} \mbox{\boldmath$\tau$}}P(\mbox{\boldmath$\sigma$})
    P(\mbox{\boldmath$\tau$})\left\langle{\mathrm e}^{\frac{1}{c}\beta
        J \sum_\alpha\sigma_\alpha\tau_\alpha}-1\right\rangle_J\right]
  \,
  \label{freen2}
\end{align}
over the densities $\{P(\mbox{\boldmath$\sigma$}),
\hat{P}(\mbox{\boldmath$\sigma$})\}$ for the free energy per site.
The saddle-point equations become
\begin{align}
  P(\mbox{\boldmath$\sigma$})=\dfrac{{\mathrm
      e}^{-\beta D\sum_\alpha
      \sigma_\alpha^2-\hat{P}(\mbox{\boldmath$\sigma$})}}
  {\sum_{\mbox{\boldmath$\sigma$}}{\mathrm e}^{-\beta D\sum_\alpha
      \sigma_\alpha^2-\hat{P}(\mbox{\boldmath$\sigma$})}}
  \label{saddle1}
\end{align}
and
\begin{align}
  \hat{P}(\mbox{\boldmath$\sigma$})=-c\sum_{\mbox{\boldmath$\tau$}}
  P(\mbox{\boldmath$\tau$})\left\langle{\mathrm e}^{\frac{1}{c}\beta J
      \sum_\alpha\sigma_\alpha\tau_\alpha}-1\right\rangle_J\,.
  \label{saddle2}
\end{align}
Eliminating $\hat{P}(\mbox{\boldmath$\sigma$})$ in Eq.
(\ref{saddle1}) by means of this expression we obtain the
self-consistency relationship
\begin{align}
  P(\mbox{\boldmath$\sigma$})=\dfrac{\exp\left(-\beta D\sum_\alpha
      \sigma_\alpha^2+c\sum_{\mbox{\boldmath$\tau$}}
      P(\mbox{\boldmath$\tau$})\left\langle{\mathrm
          e}^{\frac{1}{c}\beta J \sum_\alpha\sigma_\alpha\tau_\alpha}
        -1\right\rangle_J\right)}
  {\sum_{\mbox{\boldmath$\sigma$}}\exp\left(-\beta D\sum_\alpha
      \sigma_\alpha^2+c\sum_{\mbox{\boldmath$\tau$}}
      P(\mbox{\boldmath$\tau$})\left\langle{\mathrm
          e}^{\frac{1}{c}\beta J \sum_\alpha\sigma_\alpha\tau_\alpha}
        -1\right\rangle_J\right)}\,
  \label{saddle3}
\end{align}
and the free energy becomes
\begin{align}
  f(\beta)=-\lim_{n\rightarrow 0}&\frac{1}{\beta n}
  \left[-\frac{c}{2}\sum_{\mbox{\boldmath$\sigma$}
      \mbox{\boldmath$\tau$}}P(\mbox{\boldmath$\sigma$})
    P(\mbox{\boldmath$\tau$})\left\langle{\mathrm e}^{\frac{1}{c}\beta
        J\sum_\alpha\sigma_\alpha\tau_\alpha}-1\right\rangle_J\right.\\
  &\left.+\log\sum_{\mbox{\boldmath$\sigma$}}\exp\left(-\beta D\sum_\alpha
      \sigma_\alpha^2+c\sum_{\mbox{\boldmath$\tau$}}
      P(\mbox{\boldmath$\tau$})\left\langle{\mathrm
          e}^{\frac{1}{c}\beta J\sum_\alpha\sigma_\alpha\tau_\alpha}
        -1\right\rangle_J\right)\right]^{*} \,.\nonumber
  \label{freen3}
\end{align}
with the order function given by the saddle-point equation, indicated
by the asterisk.

Our search for solutions of Eq.(\ref{saddle3}) will be restricted to
the replica symmetry (RS) ansatz, \cite{KS87,Mo98}. This means that
$P(\mbox{\boldmath$\sigma$})$ should remain invariant under replica
permutations and, consequently, for three-state spins it should only
depend on the summations $\sum_\alpha\sigma_\alpha$ and
$\sum_\alpha\left(\sigma_\alpha\right)^2$, with their corresponding
weights. Thus, in extension of the RS ansatz for the order function
in finite-connectivity two-state Ising models \cite{NC04}, we assume
that for the three-state model
\begin{align}
  P(\mbox{\boldmath$\sigma$})=\int dh\,db\,W(h,b) \dfrac{{\mathrm
      e}^{\beta h\sum_\alpha\sigma_\alpha -\beta
      b\sum_\alpha\sigma_\alpha^2}}{\left[2{\mathrm e}^{-\beta
        b}\cosh(\beta h)+1\right]^n}\,\,\,,
\end{align}
for any real $n$, where $h$ and $b$ are the two components of the
local field and $W(h,b)$ is a density which has to be determined
self-consistently. Since the normalization factors, both here and in
Eq.(\ref{saddle3}), become one in the limit $n\rightarrow 0$ we may
leave them aside. Expanding the second exponential in
Eq.(\ref{saddle3}) and using the RS ansatz we have
\begin{align}
  \nonumber P(\mbox{\boldmath$\sigma$})={\mathrm e}^{-\beta D
    \sum_\alpha\sigma_\alpha^2}\sum_{k=0}^\infty\frac{{\mathrm
      e}^{-c}c^k}{k!}&\int\prod_{l=1}^k
  dh_l\,db_l\,W\left(h_l,b_l\right)
  dJ_l\,p\left(J_l\right)\\
  &\prod_{l=1}^k\exp\left[\sum_\alpha \log\sum_{\tau_\alpha^l}{\mathrm
      e}^{\beta h_l\tau_\alpha^l-\beta
      b_l\left(\tau_\alpha^l\right)^2+\frac{1}{c}\beta
      J_l\sigma_\alpha\tau_\alpha^l}\right]\,\,,
  \label{saddle5}
\end{align}
where $p\left(J\right)$ is the probability distribution for the
interaction, assumed to be continuous. Summing over $\tau_\alpha^l$
and using the identity $\sum_{\sigma}\delta_{\sigma\sigma_\alpha}=1$,
in order to extract the appropriate dependence on
$\sum_\alpha\sigma_\alpha$ and
$\sum_\alpha\left(\sigma_\alpha\right)^2$, we obtain
\begin{align}
  \nonumber P(\mbox{\boldmath$\sigma$})={\mathrm e}^{-\beta D
    \sum_\alpha\sigma_\alpha^2}&\sum_{k=0}^\infty\frac{{\mathrm
      e}^{-c}c^k}{k!}\int\prod_{l=1}^k
  dh_l\,db_l\,W\left(h_l,b_l\right)
  dJ_l\,p\left(J_l\right)\\
  &\prod_{l=1}^k\exp\sum_\alpha\sum_\sigma\delta_{\sigma\sigma_\alpha}
  \log\left[2\cosh\left(\beta h_l+\beta J_l\sigma/c\right) {\mathrm
      e}^{-\beta b_l}+1\right]\,.
  \label{saddle6}
\end{align}
For three-state spins we use the representation
\begin{align}
  \delta_{\sigma\sigma_\alpha}=1-\sigma^2-\sigma_\alpha^2+
  \sigma\sigma_\alpha/2+3\sigma^2\sigma_\alpha^2/2\,
  \label{delta}
\end{align}
to do the sum over $\sigma$ and with the RS ansatz in the left hand
side of the equation we obtain the self-consistency relationship for
$W(h,b)$,
\begin{align}
  W(h,b)=\sum_{k=0}^{\infty}\frac{{\mathrm e}^{-c}c^{k}}{k!}
  \int&\prod_{l=1}^kdh_l\,db_l\,W(h_l,b_l)dJ_lp(J_l)\\
  &\times\delta\left(h-\frac{1}{\beta}\sum_{l=1}^k\phi\left(h_l,b_l,J_l\right)
  \right)\delta\left(b-D+\frac{1}{\beta}\sum_{l=1}^k
    \psi\left(h_l,b_l,J_l\right) \right)\,\,\,,
  \label{distr}
\end{align}
where
\begin{align}
  \phi\left(h,b,J\right)=\frac{1}{2}\log\frac{2a_{1}+{\mathrm
      e}^{\beta b}} {2a_{-1}+{\mathrm e}^{\beta b}}
  \label{phi}
\end{align}
and
\begin{align}
  \psi\left(h,b,J\right)=\frac{1}{2}\log\frac{(2a_{1}+{\mathrm
      e}^{\beta b})(2a_{-1} +{\mathrm e}^{\beta b})}
  {\left[
      2a_{0}+{\mathrm e}^{\beta b}\right]^2}\,\,\,,
  \label{psi}
\end{align}
in which $a_{\sigma}=\cosh(\beta h+\sigma\beta J/c)$\,.

To determine the density $W\left(h,b\right)$ we proceed numerically
by means of population dynamics of a large number of fields updated
as follows \cite{MP01}, for each value of $\beta$ and $D$.
First a number $k$ is chosen from a Poisson distribution of mean
$c$. Then, cells $\left(h_l,b_l\right)$ and couplings $J_l$ with $l$
running from 1 to $k$ are selected at random from the population and
the summations in the delta functions are calculated. Next, one
selects at random a new cell $(h,b)$ from the population and sets
\begin{align}
  h=\frac{1}{\beta}\sum_{l=1}^k\phi\left(h_l,b_l,J_l\right)\,\,,
  \label{distrh}
\end{align}
\begin{align}
  b=D-\frac{1}{\beta}\sum_{l=1}^k
  \psi\left(h_l,b_l,J_l\right)\,\,
  \label{distrb}
\end{align}
continuing the procedure until it converges to a limiting $W(h,b)$.

Knowledge of $W(h,b)$ allows to determine the magnetization
\begin{align}
  m=\int\,dh\,db\,W(h,b)\left\langle\sigma\right\rangle\,,
  \label{m}
\end{align}
the spin-glass order parameter
\begin{align}
  q=\int\,dh\,db\,W(h,b)\left\langle\sigma\right\rangle^2
  \label{q}
\end{align}
and the additional parameter
\begin{align}
  r=\int\,dh\,db\,W(h,b)\left\langle\sigma^2\right\rangle\,,
  \label{r}
\end{align}
for the three-state model, where $\left\langle\sigma\right\rangle$
and $\left\langle\sigma^2\right\rangle$ depend on $h$ and $b$ as
\begin{align}
  \left\langle\sigma\right\rangle=\frac{\sinh(\beta h)}{\cosh(\beta
    h)+{\mathrm e}^{\beta b}/2}
  \label{sigma}
\end{align}
and
\begin{align}
  \left\langle\sigma^2\right\rangle=\frac{\cosh(\beta h)}{\cosh(\beta
    h)+{\mathrm e}^{\beta b}/2}\,.
  \label{sigma2}
\end{align}

As usual, $m\neq 0$ indicates magnetic ordering, while $m=0$ and
$q>0$ corresponds to spin-glass ordering. The additional parameter
$r$ is zero only when all spins are in the $\sigma_i=0$ local state.

The free energy has energetic and entropic contributions given by
\begin{align}
  \nonumber f_1(\beta)&=\frac{c}{2\beta}\int
  dh\,dh'\,db\,db'\,dJ\,P(J)W(h,b)W(h',b')\\
  \nonumber
  &\quad\quad\quad\quad\quad\times\left\{\log\left[1/4+{\mathrm
        e}^{-\beta (b+b')}\Bigl(\cosh(\beta
      h)\cosh(\beta h')\cosh(\beta J/c) \right.\right.\Bigr.\\
  &\Bigl.\left.\left.+\sinh(\beta h)\sinh(\beta h')\sinh(\beta
      J/c)\Bigl)+{\mathrm e}^{-\beta b}\cosh(\beta h)/2+{\mathrm
        e}^{-\beta b'}\cosh(\beta h')/2\right]\right.\\
  \nonumber & \quad\quad\quad\quad\quad\quad\quad\quad
  \left.-\log\left[\Bigl({\mathrm e}^{-\beta b}\cosh(\beta
      h)+1/2\Bigr)\left({\mathrm e}^{-\beta b'}\cosh(\beta
        h')+1/2\right)\right]\right\}
\end{align}
and
\begin{align}
  \nonumber f_2(\beta)=-\frac{1}{\beta}&\sum_{k=0}^\infty
  \frac{c^k{\mathrm e}^{-c}}{k!}\int\prod_\ell
  dh_\ell\,db_\ell\,dJ_\ell\,W(h_\ell,b_\ell)P(J_\ell)\\
  &\times \log\biggl[1+2\exp\biggl(\sum_\ell
  \psi\left(h_\ell,b_\ell,J_\ell\right)-\beta D\biggr)\cosh\biggl(\sum_\ell
  \phi\left(h_\ell,b_\ell,J_\ell\right)\biggr)\biggr]\,,
  \label{entrop}
\end{align}
respectively, within the RS ansatz.
\section*{Results}

\begin{figure}
  \begin{tabular}{c}
    \includegraphics[width=8cm,clip]{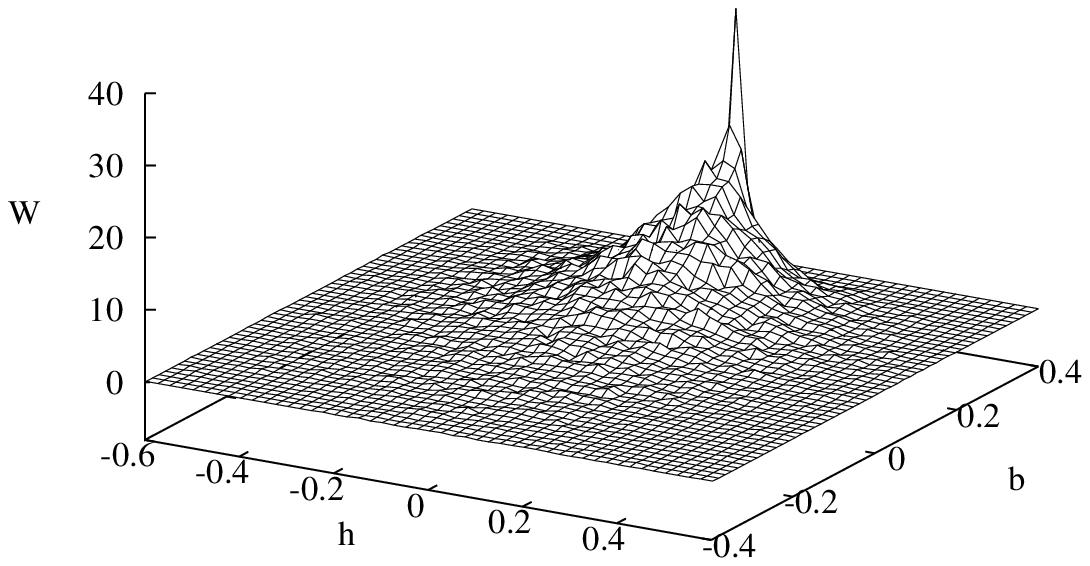}\\
    (a)\\
  \end{tabular}
  \begin{tabular}{c}
    \includegraphics[width=8cm,clip]{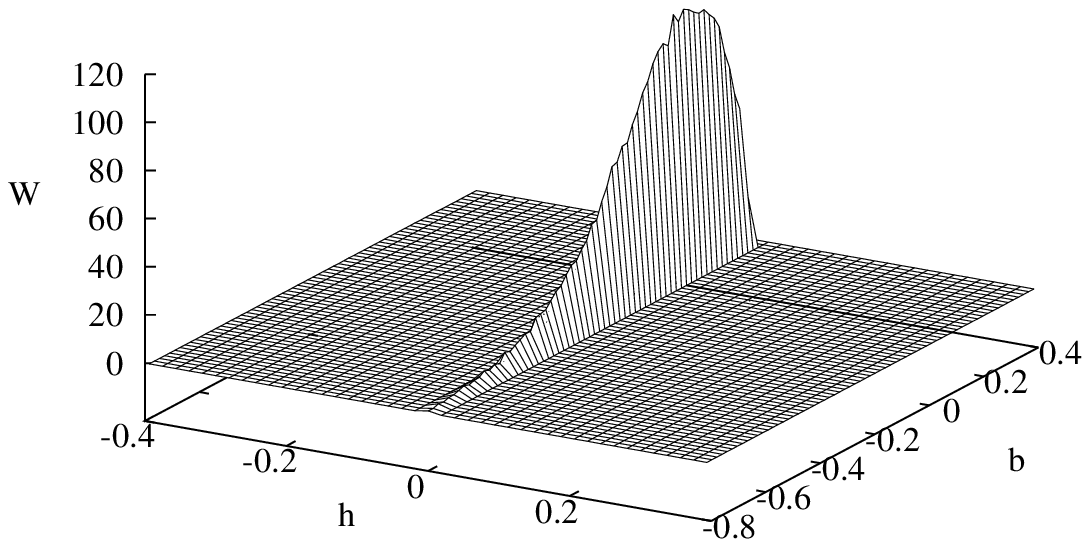}\\
    (b)\\
  \end{tabular}

  \caption{Local field distribution $W(h,b)$ for $c=6.0$,
    $D=0.35$ and $T=0.1$, with initial conditions (A) and (B)
    specified in the text in (a) and (b), respectively.}
  \label{dist}
\end{figure}

In what follows we assume that the couplings $J_{ij}$ are
independent Gaussian random variables with zero mean and unit
variance. In the binary Ising SG model, there is a one component
local field $h$ and a density $W(h)=\delta(h)$ in the P phase. The
critical temperature for a bifurcation to a SG solution can be found
in that case by means of an expansion of the full $W(h)$ around
$\delta(h)$. (See ref. \cite{NC04} for details.) In the case of the
present three-state SG model with finite connectivity, there is no
simple form for the density of the two-component local field
$W(h{,}b)$, even in the P phase, and in order to obtain the
thermodynamic properties one has to use a population dynamics
procedure to calculate explicitly that density. Since the mean of
$J_{ij}$ is zero, there is no long-range order and we expect that
$W(h{,}b)$ is symmetrically distributed around the $h=0$ axis.
Depending on the values of $r$ and $q$, the system can be found
either in a SG or in a P phase.

The implementation of population dynamics requires an initial guess
for $W(h{,}b)$, to be constructed as follows. A finite surface on
the space $(h{,}b)$ is divided into $n\times n$ cells and this set
of $n^2$ cells is populated with $N$ vector fields. In this work,
the initial population was distributed in two distinct ways: (A) the
two components of each field were randomly distributed between -0.5
and 0.5 and (B) all the fields had $h=0$ and $b$ randomly
distributed between $D-1$ and $D$. For each set of
parameters $c$, $D$ and $T$ the population dynamics is allowed
to run until a stationary distribution $W(h{,}b)$ is reached.

Illustrative examples of stationary distributions for $c=6$, $D=0.35$
and $T=0.1$, computed with $n=128$ and $N=40\,000$ are shown in
Fig. \ref{dist}. Starting from the initial condition (A), the
population dynamics converges to the two-dimensional distribution,
symmetric around $h=0$, shown in Fig. \ref{dist} (a). This field
distribution leads to a spin-glass solution with $r>0$ and $q>0$,
according to Eqs. (\ref{r}) and (\ref{q}). Starting with the initial
condition (B), $W(0{,}b)$ converges to the one-dimensional
distribution shown in Fig. \ref{dist} (b). Here, only fields along
the $h=0$ axis remain populated, and there is a long tail along the
$b$ axis. This field distribution leads to a paramagnetic phase with
$r>0$ and $q=0$. Thus, for the chosen set of parameters, the initial
conditions (A) and (B) are within the basins of attraction of the SG
and P solutions, respectively.

\begin{figure}
  \includegraphics[width=8cm,clip]{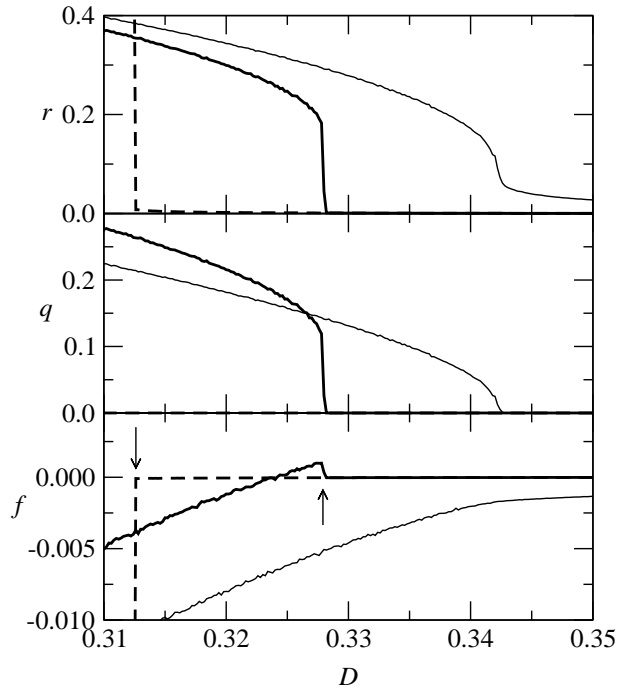}\quad\quad

  \caption{Order parameters $r$ and $q$ and free energy per site
  $f$ as functions of $D$, for $c=8.0$, exhibiting a continuous
  transition at $T=0.07$ (light solid lines) and a first-order
  transition at $T=0.04$ (heavy solid lines for SG states and dashed
  lines for P states) discussed in the text. The arrows indicate
  points on the spinodals.}
  \label{rq1}
\end{figure}

Specific examples of results for the dependence of the order
parameters with $D$ near the transition between the SG and the P
phase obtained from Eqs. (\ref{q}) and (\ref{r}), that were used to
construct the phase boundaries are shown in Fig. \ref{rq1} for
connectivity $c=8.0$ and two typical situations, one of the higher
and the other one of the lower temperature behavior in which
$T=0.07$ and $T=0.04$, respectively. The corresponding free energy
for both situations, obtained by means of Eqs. (30) and
(\ref{entrop}), is shown in the lower panel of that figure. Each
point on the curves is a result of an average over ten runs of the
population dynamics procedure, in order to smooth out fluctuations.
In the presence of multiple solutions for the parameters, we follow
previous works on the fully connected model and choose the largest
one in magnitude, that yields the lowest free energy which can be
smoothly continued from one phase to the other with a change in $D$.
The curves in Fig. \ref{rq1} are rather close to those obtained
already with a lower resolution, say $n=64$ and $N=10\,000$,
indicating that the dynamics is  near convergence. To test this
point we found that further results for $r$, $q$ and $f$ obtained
with $n=256$ and $N=160\,000$ at $T=0.04$, which is the case where
the transition is first-order, are almost indistinguishable from
those in Fig. \ref{rq1}.

As $D$ increases for $T=0.07$, $q$ and $r$ decrease continuously and
the transition from the SG to the P phase takes place at $q=0$ and
small $r$. When $T=0.04$, instead, there is a genuine first-order
transition signalled by a continuous free energy with a
discontinuity of the entropy and the appearance of a pair of
spinodals, one for states (in solid lines) attained from the
spin-glass phase and the other one for states (in dashed lines)
reached from the paramagnetic phase with increasing or decreasing
values of $D$, respectively. The arrows indicate the points on the
spinodals for that value of $T$, and the transition between the
phases appears at the crossing of the free energies, where
$D=0.328$. The first-order transition persists for somewhat higher
$T$ with the spinodals becoming closer up to the merging with the
continuous transition at a tricritical point given by $T\simeq
0.059$ and $D\simeq 0.33$ for $c=8.0$, and similar results are
obtained for other values of $c\geq c^{*}$ where $c^{*}$ is between
$5.6$ and $6.0$, whereas only a continuous transition appears for
smaller $c$ at all $T$. It is interesting to note that the
changeover from a continuous to a discontinuous transition at low
$T$ is preceded by a discontinuous transition between two SG states
{\it{within}} the SG phase for $c^{*}\simeq 5.6$. For $c$ slightly
larger this transition merges with the continuous transition that
separates the two phases.

\begin{figure}
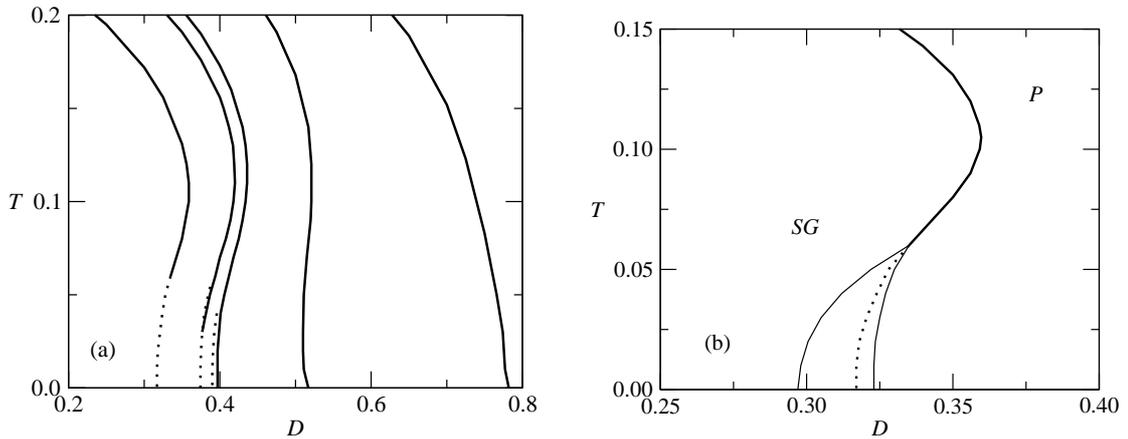

  \centerline{
    \includegraphics[width=7cm,clip]{2-8_.eps}\quad\quad
    \includegraphics[width=7cm,clip]{8.eps}
  }
  \caption{(a) The $T$ vs. $D$ phase diagram for connectivity
    $c=2.0$, $4.0$, $5.6$, $6.0$ and $8.0$ from right to left. Solid
    (dotted) lines denote continuous (discontinuous) transitions
    between a spin glass (SG) phase at left and a paramagnetic (P)
    phase at right of each curve. (b) Detail of the phase boundary for
    $c=8$ including the spinodals of the P and SG phases on the left
    and right of the transition in light solid lines, respectively.}
  \label{diag1}
\end{figure}

A global picture of the transitions for finite connectivity is given
by the $T$ vs. $D$ phase diagram for several values of $c\leq 8.0$
shown in Fig. \ref{diag1}. The transition is continuous for low $c$
and all $T$ with the order parameters going continuously to zero
with increasing values of $D$, and there is a reentrance for
$c\gtrsim 3.5$. On the other hand, a first-order transition appears
with increasing $c$ between $5.6$ and $6.0$ at low but finite $T$
that starts at a tricritical point on the reentrance of the
continuous phase boundary. The reentrance separates the P phase at
low $T$ from the SG phase at higher $T$. Note that the tricritical
point appears at larger values of $T$ with increasing $c$, and this
has been checked by further calculations for $c=16$. We also found
that, for all the values of $c$ larger than $\sim 2.5$ considered in
this work, the phase boundaries move to the left with increasing $c$
towards lower values of $D$. Indeed, there is a monotonic decrease
of $T_0(c)$, which is the critical $T$ at $D=0$, with increasing $c$
and we found numerically that $T_0\sim J/\sqrt{c}$ for large $c$,
which is the same behavior as that in the two-state spin-glass model
with finite connectivity where the transition is always continuous
\cite{NC04}.

\begin{figure}
  \includegraphics[width=8cm,clip]{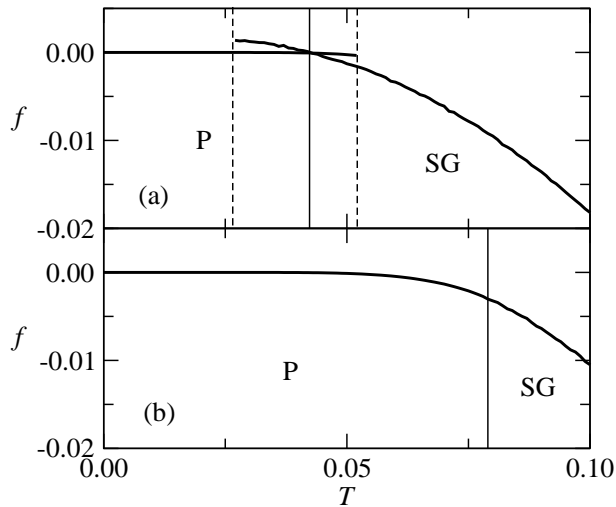}\quad\quad

  \caption{Temperature dependence of the free energy for $c=8$
  across the first-order transition at $D=0.325$ in (a) and the
  continuous transition at $D=0.35$ in (b). The dashed left
  (right) line indicates a point on the SG (P) spinodals, respectively,
  and the vertical solid lines indicate the transition with an increase
  in entropy from the P to the SG phase in (a) and with no increase in
  (b). The normal transition between the SG and the P phase, not shown
  in the figure, appears at higher $T$.}
  \label{ft1}
\end{figure}

The reentrance of the phase boundaries is a manifestation of inverse
freezing which appears either as a genuine first-order transition
with a discontinuity of the entropy or as a continuous transition
with only a gradual change in the entropy from one phase to the
other. In order to show that the discontinuity of the entropy across
the first-order transition in the GS model with finite connectivity
goes in the right direction to account for inverse freezing we show
in Fig. \ref{ft1} (a) the temperature dependence of the free energy
for $c=8$ and $D=0.325$. The free energy is continuous and
follows the lower curve when there are two solutions and, indeed,
the entropy of the P phase below the first-order transition (which
is slightly positive for that value of $D$) is smaller than the
entropy of the higher temperature SG phase, with a discontinuity of
the entropy at the transition. This means that in heating the system
the disordered paramagnet "freezes" into the amorphous SG with an
exchange of latent heat. For comparison, we show in Fig. \ref{ft1}
(b) the free energy, for $D=0.35$, with no discontinuity in the
entropy across the continuous transition.

The phase diagram for finite $c$ obtained in the present work within
the RS scheme differs from the $T$ vs. $D$ phase diagram for
the fully connected GS model in that, apparently, there is no
reentrance in the latter within that scheme but there is a
reentrance within full replica symmetry breaking. The first-order
transition line appears in that case all along the reentrance of the
phase boundary down to $T=0$ and it is even slightly continued above
into the normal phase boundary. Thus, there is no continuous
transition on the reentrance in that case. On the other hand there
is, apparently, a further reentrance on the first-order transition
at low $T$ \cite{CL05}.

The kind of behavior we find here for finite $c$ is similar to the
phase diagram of a somewhat different model of a three-state Ising
spin glass with full connectivity in which the degeneracy of the
active spins ($\sigma=1$ or $\sigma=-1$) is larger than the
degeneracy of the inactive state ($\sigma=0$), in contrast to the GS
model where these degeneracies are the same. Indeed, the $T$ vs. $D$
phase boundary in that case also displays both a first-order
transition at low $T$ and a continuous transition at higher $T$ on
the reentrance of the phase boundary, even within the RS scheme
\cite{SS03}.

\section*{Summary and conclusions}

The statistical mechanics of the binary Ising SG model with finite
random connectivity, within mean-field theory, has been extended in
this work to study the three-state Ising SG model with crystal-field
effects of Ghatak and Sherrington, which is a random-bond
Blume-Capel model. Renewed interest in this model is due to the
recent discovery of numerous physical realizations within
condensed-matter physics that exhibit inverse transitions (see ref.
\cite{PL10} for a recent review), and finite connectivity could be
of use in order to study the behavior of such systems with effective
interactions between infinite and short range or between infinite
dimensions and three-dimensions.

The replica method for disordered spin systems with finite
connectivity has been extended here within the replica symmetry (RS)
ansatz. The RS ansatz for the order function that describes the
behavior in the present model with finite connectivity introduces a
distribution of a two-component effective field which plays a
crucial role in determining the relevant order parameters of the
model. An iterative self-consistency relation is derived for the
local field distribution, an analytic solution of which does not
seem possible in the presence of a first-order transition, and we
use instead a population dynamics numerical procedure. Phase
diagrams were then obtained in order to investigate the effects of
finite connectivity and anisotropy parameter on the transitions
between a SG and a P phase, with emphasis on reentrance behavior.
The latter is a necessary feature in order to describe inverse
transitions, in particular inverse freezing, in view of the two
phases favored by the present model.

The main result of this paper is the $T$ vs. $D$ phase diagram,
for various values of the connectivity, which exhibits the different
phase transitions that may appear at low temperature. For low
connectivity the transition is a continuous one down to $T=0$, with
a reentrance at low $T$ for higher connectivity exhibiting both a
continuous and a discontinuous transition. We showed that the
discontinuous transition is a first-order transition with a
discontinuity of the entropy at the transition in which the entropy
of the SG phase is larger than the entropy of the P phase below the
transition, which is a sign of inverse freezing in the system.

The results obtained so far within the RS ansatz could be changed by
replica-symmetry breaking (RSB) and an extension in this direction
could be interesting \cite{MP01} although not so easy to carry out
in the present case due to the time consuming evaluation of the
distribution of the two-dimensional effective field. However, if the
trend of the results for the GS model with full connectivity are
taken as a guide, one would expect an enhancement of the reentrance
on the first-order transition for finite connectivity within RSB,
without much change of the continuous transition.

It may also be interesting to find out to which extent the specific
form of the distribution of random bonds makes a difference. In place
of a Gaussian with mean zero and unit variance that we used in this
work, one could have a Gaussian with mean $J_0 > 0$. We found that the
phase diagram is somewhat changed in that case. Instead, one could
consider a general bimodal distribution between ferromagnetic and
antiferromagnetic interactions which allows to predict when the
effects of frustration may become more relevant before going into a
RSB calculation, an argument that has been used before
\cite{NC04}. This, and related issues, are currently being studied.

The procedure extended in this work can be applied to study the
effects of finite connectivity in other multi-state spin models like
the random BEGC model or the degenerate spin model of Schupper and
Shnerb either with random or uniform interactions \cite {SS03,CL05}.
It can also be applied to attractor neural networks and the study of
the retrieval performance in a three-state network with finite
connectivity and a Hebbian learning rule is currently being
investigated \cite{BET10}.

\section*{Acknowledgements}

Discussions with Desir\'e Boll\'e, Paulo R. Krebs and Sergio Garcia
Magalh\~aes are gratefully acknowledged. The present work was
supported, in part, by Conselho Nacional de Desenvolvimento Científico
e Tecnológico (CNPq) and Fundação de Amparo à Pesquisa do Estado do
Rio Grande do Sul (FAPERGS), Brazil.

\end{document}